\documentclass[11pt,a4paper]{article}
\usepackage{jheppub}
\usepackage[T1]{fontenc}
\usepackage{amssymb}
\usepackage{mathtools}
\usepackage{amsmath}
\usepackage{bm}
\usepackage{stackengine}
\usepackage{scalerel}
\usepackage{framed}
\usepackage{mathtools}
\usepackage{extarrows}
\usepackage{color}
\usepackage{bbold}
\usepackage{bm}
\usepackage{latexsym}
\usepackage{braket}
\usepackage{slashed}
\numberwithin{equation}{section}
\usepackage{graphicx}
\usepackage{float} 
\usepackage{graphicx}
\usepackage{color} 
\usepackage{url}
\usepackage{tcolorbox}
\tcbuselibrary{fitting,raster,skins,xparse}
\usepackage{caption}
\usepackage{subcaption}

\usepackage{physics}
\usepackage{tikz}
\usepackage{mathdots}
\usepackage{yhmath}
\usepackage{cancel}
\usepackage{siunitx}
\usepackage{array}
\usepackage{multirow}
\usepackage{gensymb}
\usepackage{tabularx}
\usepackage{extarrows}
\usepackage{booktabs}
\usetikzlibrary{fadings}
\usetikzlibrary{patterns}
\usetikzlibrary{shadows.blur}
\usetikzlibrary{shapes}
\usepackage{algpseudocode}

\makeatletter
\newcommand{\doublewidetilde}[1]{{%
  \mathpalette\double@widetilde{#1}%
}}
\newcommand{\double@widetilde}[2]{%
  \sbox\z@{$\m@th#1\widetilde{#2}$}%
  \ht\z@=.9\ht\z@
  \widetilde{\box\z@}%
}
\makeatother

\title{\boldmath The integrable Bullough-Dodd model under celestial holography\unboldmath}

\author{Minjia Wang}
\author[1]{and Wei Fan~\note{Corresponding author. (The administrative policy of our university requires us to put students' name in the first place, so we did not follow the convention of arranging the names in the alphabet order. We apologize for this. ) }}
\affiliation{Department of Physics, School of Science, Jiangsu University of Science and Technology,
Zhenjiang, 212100, China}

\emailAdd{fanwei@just.edu.cn}

\abstract{
We study celestial amplitudes for the  S-matrix of the 2d  integrable Bullough-Dodd model.  This model has bound states that appear  as  poles in the physics strip of its 2d S-matrix, which complicates the computation of celestial amplitudes. However, it turns out that the celestial amplitudes are, in fact, well-structured. The celestial bootstrap (arising  from the unitarity and crossing symmetry of 2d S-matrix) can be decomposed into a finite-dimensional linear space, whose base-integrals evaluate into harmonic numbers. This clean structure replaces the complicated integration with simple algebra of elementary functions, and the celestial bootstrap reduces to a programmable recursion process of simple algebra. Interestingly, this linear space has a subspace that happens to cover the celestial bootstrap of the Sinh-Gordon model studied by 2209.~02776.  So  the celestial dual of these 2d  integrable models turns out to be 'bootstrapable'  in the practical sense, that is, a programmable recursion process. 
}

\begin{document}

\maketitle
\flushbottom

\section{Introduction}
\label{sec:intro}

Celestial holography  is a flat holography~\cite{Strominger:2017zoo,Pasterski:2021raf, Pasterski:2021rjz,Raclariu:2021zjz} of Minkowski spacetime.  The bulk QFT in Minkowski spacetime $\mathbb{R}^{1,3}$ can be recast as a boundary CFT in the celestial sphere. The amplitudes are converted from the  momentum basis to the  boost basis (the so-called conformal primary wavefunction)~\cite{Pasterski:2017kqt} and then become the CFT correlation functions. 

Recently, celestial holography is applied to $2d$ integrable QFTs~\cite{Duary:2022onm,Kapec:2022xjw,Stolbova:2023smk,Stolbova:2023cof}, where the dual theory is a $0d$ CFT without position dependence. The $2d$ integrable models have a well-structured exact S-matrix,  so its celestial dual provides a platform for illustrating the full quantum effects in celestial holography. In this case  the conformal primary wave function reduces to the Fourier transform, so the celestial amplitudes become Fourier transforms of 2d S-matrix. The unitarity and crossing symmetry of 2d S-matrix translate into bootstrap relations among celestial amplitudes, called celestial bootstrap, where the convolution integral of celestial amplitudes plays an important role. 

In the original work~\cite{Duary:2022onm} the Sinh-Gordon model is studied, which has no bound state. The exact S-matrix is decomposed into a perturbative series whose Fourier transform gives the perturbative  celestial amplitudes. Explicitly the  celestial amplitudes are computed for the first few orders and the convolution integral is computed at the first order.  But it is not clear whether it can be generalized to arbitrary orders in a practical sense, because the general structure of convolution integrals is not clear.  
A different approach is taken in~\cite{Kapec:2022xjw} when studying the Affine Toda theory, where the full S-matrix is used instead of decomposing into a perturbative series. But a dress term is introduced by coupling the 2d theory to gravity (also discussed in~\cite{Duary:2022onm}) to regulate the Fourier transform, because the full S-matrix does not vanish at infinity. Another approach is used  in~\cite{Stolbova:2023cof} for the $O(n)$ nonlinear sigma model. Because its 2d S-matrix is given by the Gamma functions~\cite{Zamolodchikov:1978xm},  the Fourier transform for celestial amplitudes is converted to the Mellin transform,  which finally evaluates into the Mejer G-functions. But the unitarity condition, that is the convolution integral, can not be verified explicitly in this case.  

In this paper, we study the 2d integrable Bullough-Dodd model using the perturbative approach and find that its celestial bootstrap turns out to be workable because its convolution integrals of celestial amplitudes are well-structured. The Bullough-Dodd model is an important model in the field of integrability. This model first emerges in the context of pure mathematics~\cite{tzitzeica1910nouvelle}.  Its infinity symmetries is discovered in~\cite{Dodd:1977bi} and its S-matrix is proposed in~\cite{Arinshtein:1979pb} by the methods of integrability. Its spectrum consists of one particle which is the bound state of itself. Here we take the approach of decomposing its exact S-matrix into a perturbative series, because we are interested in its celestial dual as it is. The bound state leads to poles in its 2d S-matrix. Nevertheless, the celestial amplitudes is computable from the Cauchy integration formula by summing over residues. The bootstrap relations also receive extra contributions of this bound state. However, all convolution integrals are well structured with totally 11 different kinds of basis integrals. These basis integrals are computed by the Feynman tricks of integration and  are connected with harmonic numbers. In this way, the celestial bootstrap becomes programable recursion relations. 

As a corollary, its celestial amplitudes constitute a finite-dimensional linear space of dimension-seven, where the convolution integral reduces to  basis transformation. So the celestial bootstrap becomes a very practical process: once the first order is given, celestial amplitudes at arbitrary order can be computed quickly. We have computed celestial amplitudes up to the fourth order by the direct Fourier transform, and they agree with the results of the bootstrap approach. In the direct computation of Fourier transform, higher order celestial amplitudes take increasing amount of time, because the order of pole is increasing and the sum of residues is to infinity. In the bootstrap approach, the higher order amplitudes are very efficient to compute because only basis-transformation is involved. Although we did it in a semi-automatic manner, the programming experts can do it in a full automatic manner. 

In addition,  using the above method we briefly revisit the Sinh-Gordon model studied in the original work~\cite{Duary:2022onm}. We find that its convolution integrals are also well structured with 3 of the 11 kinds of basis integrals mentioned above. And its celestial amplitudes constitute a linear subspace of dimension-three, which is a subspace of the linear space expanded by celestial amplitudes of the Bullough-Dodd model.

The paper is organized as follows. In Section~\ref{sec:preliminary}, we briefly review the necessary ingredients  needed for the computation. In Section~\ref{sec:amp}, we discuss the computation of celestial amplitudes for the Bullough-Dodd model. In Section~\ref{sec:base-integral}, we discuss the unitarity and crossing symmetry and the base-integral that is connected to the harmonic number. In Section~\ref{sec:bootstrap}, we discuss the celestial bootstrap of higher order amplitudes. In Section~\ref{sec:sinh-gordon}, we briefly revisit the Sinh-Gordon model.  We conclude with a discussion of open questions  in Section~\ref{sec:conclusion}.

\section{Preliminary}
~\label{sec:preliminary}

In this section we firstly review the simplification of conformal primary wave functions for 2d S-matrix. Then we briefly review the 2d S-matrix of the integrable  Bullough-Dodd model. We do not go into detailed methodology and literature of the 2d integrable models, because the focus is on the celestial bootstrap rather than the integrability. 

\textbf{[I]:} 
In celestial holography, a massive four-momentum $p_i^\mu=\epsilon_i m_i \hat{p}_i^\mu$ is parameterized by the mass of the particle $m_i$ and the hyperbolic coordinate $(y_i, \vec{z}_i)$ via the formula~\cite{Pasterski:2016qvg}
\begin{align}
  \hat{p}_i^\mu(y_i, \vec{z}_i)=\left(\frac{1+y_i^2+|z_i|^2}{2 y_i}, \frac{\vec{z}_i}{y_i},  \frac{1-y_i^2-|z_i|^2}{2 y_i}\right)
\end{align}
and  $\epsilon_i=\pm 1$ represents incoming/outgoing. A direction or  point $\vec{w}_i$ on the celestial sphere is connected with the null four-vector  by the formula~\cite{Pasterski:2017kqt} 
\begin{align}
  \hat{q}_i =   (1+w_i\bar{w}_i, w_i+\bar{w}_i,  -i(w_i-\bar{w}_i),1-w_i\bar{w}_i),
\end{align}
with  $\vec{w}_i=(\operatorname{Re}(w_i),\operatorname{Im}(w_i))$.
With these parametrization the  conformal primary wave function $\Phi_{\Delta}^{m}(X^\mu;\vec{w})$  for the $4d$ massive  scalar field is  constructed as
\begin{align}
  \Phi_{\Delta}^{m(\pm)}(X^\mu;\vec{w})&\coloneqq \int  \frac{d^3 \hat{p}}{\hat{p}^{0}} G_{\Delta}(\hat{p}(y,\vec{z}) ; \hat{q}(\vec{w})) e^{\pm i m \hat{p}(y,\vec{z})\cdot X}= \int  \frac{d^3 \hat{p}}{\hat{p}^{0}} \frac{1}{(-\hat{p} \cdot \hat{q})^{\Delta}} e^{\pm i m \hat{p}(y,\vec{z})\cdot X}\nonumber\\
  &= \int_0^\infty \frac{dy}{y^3}\int d^2z \left(\frac{y}{y^2+|\vec{z}-\vec{w}|^2}\right)^{\Delta} e^{\pm i m \hat{p}(y,\vec{z})\cdot X},
\end{align}
where $G_{\Delta}(y,\vec{z};\vec{w})$ is the bulk-to-boundary propagator with the scaling dimension $\Delta$. Generally for $d$-dimensional bulk space-time the scaling dimension $\Delta$ is defined by~\cite{Pasterski:2017kqt} 
\begin{align}
    \Delta=\frac{d-2}{2}+i\lambda, \quad \lambda\in\mathbb{R}. 
\end{align}
Then by celestial holography the 4d  n-point S-matrix $A_n(p_1,p_2,\ldots,p_n) $ is dual to the celestial amplitude or n-point conformal correlation function $\mathcal{A}_n(\{\vec{w}_i\})$ as
\begin{align}
\label{eq:mellin}
\mathcal{A}_n(\{\vec{w}_i\})\coloneqq\left\langle \prod_{i=1}^n   \Phi_{\Delta_i}^{m_i}(\vec{w}_i)  \right\rangle = \left(\prod_{i=1}^n \int \frac{d^3 \hat{p}_i}{\hat{p}_i^0}G_{\Delta_i}(\hat{p}_i;\hat{q}_i)\right)   A_n(p_1,\ldots,p_n) \delta^{4}(\sum_{i=1}^n p_i),
\end{align}

If the physics happens only in one space direction, the 4d momentum is essentially reduced to 2d momentum. In this  case, the celestial sphere shrinks to a celestial point $\vec{z}=\vec{w}=0$ and the scaling dimension becomes purely imaginary $\Delta=i\lambda$.  Usually the massive 2d momentum is  parameterized by the rapidity parameter $\theta$ as $p^\mu = m \left(\cosh \theta , \sinh \theta \right)$, then the conformal primary wave function  reduces to the Fourier transform
\begin{align}
\Phi_{\Delta}^{m(\pm)}(X^\mu)=\int \frac{d \hat{p}^1}{\hat{p}^{0}}  \frac{1}{(-\hat{p} \cdot \hat{q})^{\Delta}} e^{\pm i m \hat{p}(y,\vec{z})\cdot X}= \int_{-\infty}^\infty d\theta e^{i\theta\lambda} e^{\pm i m \hat{p}(y,\vec{z})\cdot X}.
\end{align}
For a 2d S-matrix $S_{2\to2}$ of 2-to-2 scattering, the corresponding celestial amplitude, as firstly  proposed in the original work~\cite{Duary:2022onm}, simplifies to a Fourier transform 
\begin{align}
\mathcal{A}_n(\omega)\propto \delta(\Delta_+) \int_{-\infty}^\infty d\theta e^{i\theta\omega} S(\theta),\quad \Delta_{ \pm}=\frac{1}{2}\left(\Delta_1+\Delta_4 \pm \Delta_2 \pm \Delta_3\right), \theta=\theta_4-\theta_3,
\end{align}
where a redefinition  $\omega=\Delta_{-}$ is used and $S(\theta)$ is the usual 2d S-matrix of integrable models. In the following we take the convention of omitting the overall factor including the $\delta(\Delta_+)$ and focus on the Fourier transform part. 

\textbf{[II]:} The Lagrangian of the 2d Bullough-Dodd model can be written as~\cite{Alekseev:2011my}
\begin{align}
\label{eq:Largangian}
\mathcal{L}=\frac{1}{16\pi}(\partial \varphi)^2+\mu(e^{\sqrt{2}b\varphi}+2e^{-\frac{b}{\sqrt{2}}\varphi}),
\end{align}
where the exact S-matrix of 2-to-2 scattering proposed by~\cite{Arinshtein:1979pb} takes the form 
\begin{align}
\label{eq:S-matrix}
S(\theta)=\frac{\tanh\frac{1}{2}(\theta+\frac{2i\pi}{3})\tanh\frac{1}{2}(\theta-\frac{2i\pi}{3bQ})\tanh\frac{1}{2}(\theta-\frac{2i\pi b}{3Q})}{\tanh\frac{1}{2}(\theta-\frac{2i\pi}{3})\tanh\frac{1}{2}(\theta+\frac{2i\pi}{3Qb})\tanh\frac{1}{2}(\theta+\frac{2i\pi b}{3Q})}, \quad Q=b^{-1}+b.
\end{align}
In terms of the rapidity parameter, the unitarity and crossing symmetry of this S-matrix become
\begin{align}
& S(\theta)=S(i \pi-\theta) \nonumber\\
& |S(\theta)|^2=S(\theta)S(-\theta)=1.
\end{align}
The $S(\theta)$ has two poles $\theta=i \pi / 3$ and $\theta=2 \pi i / 3$ at the physical strip, which is the region $0 \leq \operatorname{Im} \theta \leq \pi$ and poles here represent the bound states. For this model, the particle appears as the bound state itself in the scattering process. 
Here we omit the details of the 2d S-matrix and refer interested readers to the review paper~\cite{Zamolodchikov:1978xm,Dorey:1996gd} for detailed methodology and literature. 

\section{Celestial amplitudes of Bullough-Dodd model}
~\label{sec:amp}

Now we evaluate the celestial dual of the Bullough-Dodd model by the perturbative approach, that is, expanding the exact S-matrix as a perturbative series

\begin{align}
\label{eq:Largangian}
S(\theta)= 1+ b^{2} S^{(1)}(\theta)+ b^{4} S^{(2)}(\theta)+\ldots,
\end{align}
and computing the celestial amplitudes order-by-order by
\begin{align}
\label{eq:amplitude_integral}
\mathcal{A}(\omega)&=\int^\infty_{-\infty}d\theta e^{i\omega\theta}S(\theta)=2\pi\Big(\delta(\omega)+b^2f_1(\omega)+b^4f_2(\omega)+\ldots\Big).
\end{align}
This perturbative 2d S-matrix $S^{(n)}(\theta)$ is equivalent to the S-matrix computed from the expanded Lagrangian order-by-order~\cite{Dorey:1996gd} 
\begin{align}
\label{eq:Series[L]}
\mathcal{L}=\frac{1}{16\pi}(\partial_{\nu}\varphi)^2+3\mu+\frac{3\mu b^2}{2}\varphi^2+\frac{\sqrt{2}\mu b^3}{4}\varphi^3+\frac{3\mu b^4}{16}\varphi^4+\cdots.
\end{align}

The perturbative expansion of $S(\theta)$ is performed  in an empirical manner. We have expanded it  to the 8th order and find that $S^{(n)}(\theta)$ satisfy the  relation  
\begin{align}
\label{eq:sn-expansion}
    S^{(n)}(\theta)\propto \frac{\mathrm{csch}^{n}(\theta)}{\left(1+2\cosh(2\theta)\right)^{n}}\Big(\text{upto }\sinh{(3n-1)\theta} \text{ or } \cosh{(3n-1)\theta}\Big) \overset{|\theta|\to\infty}{\longrightarrow} e^{-|\theta|} = 0,
\end{align}
with  three different series of n-th order poles in the complex plane at
\begin{align}
   \theta=ik\pi, \quad i (\pi / 3 + k\pi), \quad  i (2 \pi / 3 + k\pi), \quad k\in \mathbb{Z}.
\end{align}
By the spirit of integrability we intuitively can  expect that there exists a hidden closed formula and this relation is satisfied to arbitrary order,  although this closed formula of  series expansion is unknown. \textit{Nevertheless}, this relation is later confirmed in hindsight by the celestial bootstrap in Section~\ref{sec:bootstrap}  and everything appears to be consistent with each other. To keep conciseness,  the series is  only shown up to the 3rd order as follows
\begin{align}
\label{eq:Series[S]}
&S^{(1)}(\theta)=-\frac{4 i\pi\cosh(2\theta)\mathrm{csch}(\theta)}{1+2\cosh(2\theta)}\notag\\
&S^{(2)}(\theta)=\frac{2 i \pi  \text{csch}^2(\theta ) \left((9-5 \sqrt{3} \pi ) \sinh (\theta )+(\pi  \sqrt{3}+9) \sinh (5 \theta )+18 i \pi\Big(  \cosh (4 \theta )+1\Big)\right)}{9 (2 \cosh (2 \theta )+1)^2}\nonumber\\
&S^{(3)}(\theta)=\frac{\pi  \text{csch}^3(\theta ) }{27 (2 \cosh (2 \theta )+1)^3}  \Bigg(12\pi (5 \sqrt{3} \pi-9  ) \sinh (\theta )-2 (2 \sqrt{3} \pi-9  ) \sinh (3 \theta ) \nonumber\\
&+(\pi  \sqrt{3}+9 ) \sinh (7 \theta )+i \Big(-2 (12 \pi  \sqrt{3}-154 \pi ^2-27) \cosh (2 \theta ) +(30 \pi  \sqrt{3}+50 \pi ^2-27) \cosh (4 \theta )\nonumber\\
&\quad +72 \pi ^2 \cosh (6 \theta )-(6 \pi  \sqrt{3}-2 \pi ^2+27) \cosh (8 \theta )\Big)\Bigg).
\end{align}

\begin{figure}[h]
        \centering
        \begin{subfigure}[b]{0.3\textwidth}
             \centering
             \includegraphics[width=1\linewidth]{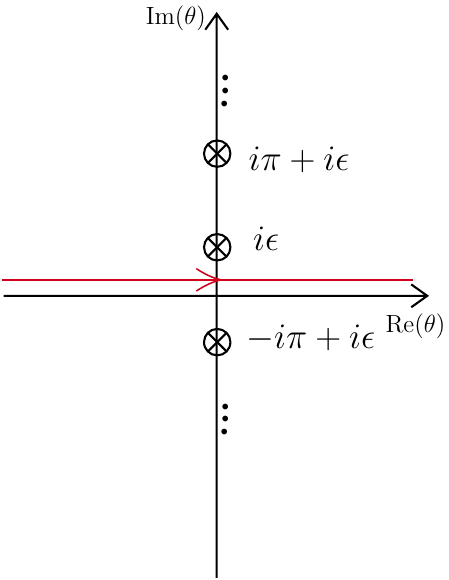} 
             \caption{$i n \pi +i\epsilon$.}
             \label{fig:pole-npi-up}
        \end{subfigure}
           \hfill
        \begin{subfigure}[b]{0.3\textwidth}
             \centering
             \includegraphics[width=1\linewidth]{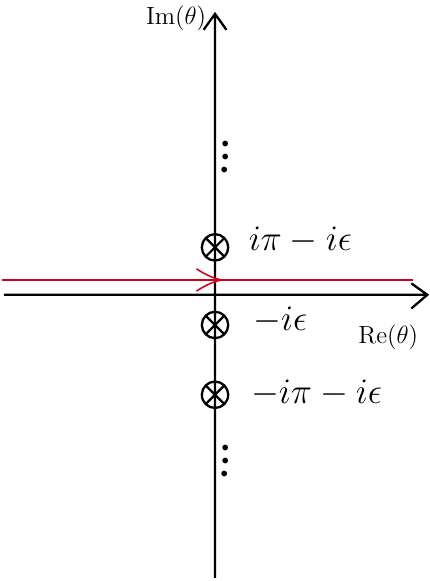} 
             \caption{$i n \pi -i\epsilon$.}
             \label{fig:pole-npi-down}
        \end{subfigure}    
           \hfill
        \begin{subfigure}[b]{0.3\textwidth}
             \centering
             \includegraphics[width=1\linewidth]{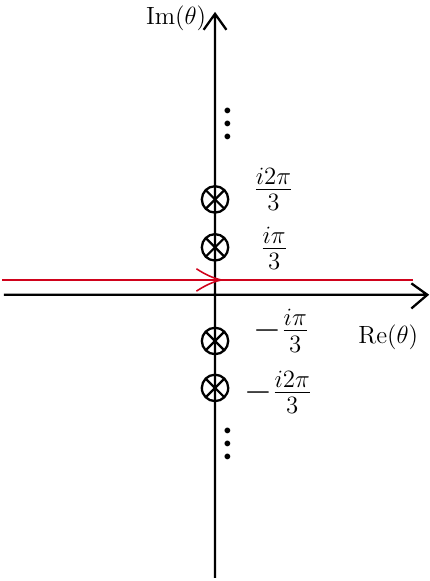} 
             \caption{$i\frac{\pi}{3}+in\pi$ and $i\frac{2\pi}{3}+in\pi$.}
             \label{fig:pole-onethird}
        \end{subfigure} 
        \caption{Poles of $\frac{ \text{csch}^n(\theta )}{ (2 \cosh (2 \theta )+1)^n}$ in $S^n(\theta)$. The regulator $\pm i\epsilon$ only affects the pole  $\theta=0$.}
        \label{fig:pole-structure}
\end{figure}

Because $S^{(n)}(\theta)$ contains a pole $\theta=0$ on the real axis, the Fourier transform is regulated by the $i\epsilon$ prescription like the retarded/advanced Green's functions in quantum mechanics and can be evaluated by the residue theorem
\begin{align}
\label{eq:def-fpm}
f_{n}^{\pm}(\omega)&=\int^\infty_{-\infty}\frac{d\theta}{2\pi} e^{i\omega\theta}S^{(n)}(\theta \pm i\epsilon) =\sum i\mathrm{Res}\left[e^{i\omega\theta}S^{(n)}(\theta)\right].
\end{align}
The pole structure is plotted in Figure~\ref{fig:pole-structure} and only the pole at $\theta=0$ is affected by the $\pm i\epsilon$ prescription. The contour in Figure~\ref{fig:pole-structure} can be closed up or down depending on the sign of $\omega$ and the encircled multiple poles contribute as an infinite sum. 

In practical computation, we use the Cauchy integration formula to evaluate the integral. At each multiple pole, this is the same as the residue of the total integrand. But the structure of $f_{n}^{\pm}(\omega)$ is exhibited more clearly in the Cauchy integration formula. As the n-th order amplitude $S^{(n)}(\theta)$ has a multiple pole of n-th order at $\theta_0$, it can be written in a meromorphic form as $g(\theta)/\left((\theta-\theta_0)^n h(\theta)\right)$ where  $h(\theta)$ has no zero. So the Fourier integral can be evaluated by the Cauchy integration formula as
\begin{align}
\oint_{\theta_0} \frac{d\theta}{2\pi} &\frac{e^{i\omega\theta}g(\theta)}{(\theta-\theta_0)^n h(\theta)} =i \frac{1}{h(\theta_0)}\Bigg(
\frac{1}{(n-1)!}\left[e^{i\omega\theta}g(\theta)\right]^{(n-1)}_{\theta_0} +\ldots \nonumber\\
&+ \Big(-\frac{h^{(n-1)}(\theta_0)}{(n-1)!h(\theta_0)} +\ldots+\left(-\frac{h^{(1)}(\theta_0)}{h(\theta_0)}\right)^{n-1}\Big)\left[e^{i\omega\theta}g(\theta)\right]_{\theta_0}
\Bigg),
\end{align}
where the superscript $(k)$ means $k$-th order derivative and this is a programmable computation. We can see that the contribution of the n-th order multiple pole $\theta_0$ is a polynomial $P_{n-1}(\omega)$ of order $(n-1)$ multiplied with an exponential factor
\begin{align}
\label{eq:sn-residue}
    \mathrm{Res}\left[e^{i\omega\theta}S^{(n)}(\theta)\right] \propto P_{n-1}(\omega) e^{i\omega\theta_0}.
\end{align}
After summing over all the poles encircled in the contour, it turns out only $7$ different exponential factors $Exp(k\pi\omega/3), k=0,1,\ldots,6$ are involved, that is, $f_{n}^{\pm}(\omega)$ constitutes a finite-dimensional linear space with the following $7$ basis 
\begin{align}
\label{eq:7basis}
   f_{n}^{+}(\omega) =\frac{1}{e^{2 \pi  \omega }-1} \sum_{k=0}^{5} P^{+}_{k;n-1}(\omega) e^{\frac{k\pi\omega}{3}}, \quad f_{n}^{-}(\omega) =\frac{1}{e^{2 \pi  \omega }-1} \sum_{k=1}^{6} P^{-}_{k;n-1}(\omega) e^{\frac{k\pi\omega}{3}},
\end{align}
which will be proved by the celestial bootstrap in Section~\ref{sec:bootstrap}.

In this way we have computed the celestial amplitudes up to the 4th order. To keep conciseness, we only show the first three orders below and $f_{4}^{\pm}(\omega)$ is moved to the appendix. The perturbative retarded celestial amplitudes are
\begin{align}
\label{eq:perturbative_celestial_amplitude_retarded}
&f^+_1=\frac{2 \pi  \left(e^{\frac{\pi  \omega }{3}}-e^{\frac{2 \pi  \omega }{3}}-2 e^{\pi  \omega }-e^{\frac{4 \pi  \omega }{3}}+e^{\frac{5 \pi  \omega }{3}}+2\right)}{3 \left(e^{2 \pi  \omega }-1\right)}\notag\\
&f^+_2=\frac{2 \pi}{9  (e^{2 \pi  \omega }-1)} \Bigg(e^{\frac{\pi  \omega }{3}} (2 \pi    \omega -3  +4 \pi \sqrt{3})+e^{\frac{2 \pi  \omega }{3}} (2 \pi   \omega +3 -4\sqrt{3} \pi )+2  e^{\pi  \omega } (2 \pi   \omega +3 )\nonumber\\
&{\quad} +\sqrt{3} e^{\frac{4 \pi  \omega }{3}}(\sqrt{3}+4 \pi ) -\sqrt{3} e^{\frac{5 \pi  \omega }{3}} (\sqrt{3}+4 \pi ) +2 (2  \pi  \omega -3 )\Bigg)\notag\\
&f^+_3=\frac{2 \pi}{27  (e^{2 \pi  \omega }-1)} \Bigg( e^{\frac{\pi  \omega }{3}} (2 \pi ^2 \omega ^2+4 \sqrt{3}\pi  (2 \pi -\sqrt{3}) \omega +23 \pi ^2-24 \sqrt{3} \pi +9)\nonumber\\
&+e^{\frac{2 \pi  \omega }{3}} (-2 \pi ^2 \omega ^2+4  \sqrt{3}\pi (2 \pi -\sqrt{3})  \omega -23 \pi ^2+24 \pi  \sqrt{3}-9)+ 2 e^{\pi  \omega } (-2 \pi ^2 \omega ^2-12 \pi  \omega +4 \pi ^2-9)\nonumber\\
&{\quad}-e^{\frac{4 \pi  \omega }{3}}(24 \pi  \sqrt{3}+23 \pi ^2+9) + e^{\frac{5 \pi  \omega }{3}}(24 \pi  \sqrt{3}+23 \pi ^2+9) +2  \left(2 \pi ^2 \omega ^2-12 \pi  \omega -4 \pi ^2+9\right)
\Bigg).
\end{align}
Similarly the perturbative advanced celestial amplitude are
\begin{align}
\label{eq:perturbative_celestial_amplitude_advanced}
&f^-_1=\frac{2 \pi  \left(e^{\frac{\pi  \omega }{3}}-e^{\frac{2 \pi  \omega }{3}}-2 e^{\pi  \omega }-e^{\frac{4 \pi  \omega }{3}}+e^{\frac{5 \pi  \omega }{3}}+2e^{2\pi  \omega }\right)}{3 \left(e^{2 \pi  \omega }-1\right)}\notag\\
&f^-_2=\frac{2 \pi}{9  (e^{2 \pi  \omega }-1)} \Bigg(e^{\frac{\pi  \omega }{3}} (2 \pi    \omega -3  +4 \pi \sqrt{3})+e^{\frac{2 \pi  \omega }{3}} (2 \pi   \omega +3 -4\sqrt{3} \pi )+2  e^{\pi  \omega } (2 \pi   \omega +3 )\nonumber\\
&{\quad} +\sqrt{3} e^{\frac{4 \pi  \omega }{3}}(\sqrt{3}+4 \pi ) -\sqrt{3} e^{\frac{5 \pi  \omega }{3}} (\sqrt{3}+4 \pi ) +2 (2  \pi  \omega -3 )e^{2 \pi  \omega }\Bigg)\notag\\
&f^-_3=\frac{2 \pi}{27  (e^{2 \pi  \omega }-1)} \Bigg( e^{\frac{\pi  \omega }{3}} (2 \pi ^2 \omega ^2+4 \sqrt{3}\pi  (2 \pi -\sqrt{3}) \omega +23 \pi ^2-24 \sqrt{3} \pi +9)\nonumber\\
&+e^{\frac{2 \pi  \omega }{3}} (-2 \pi ^2 \omega ^2+4 \sqrt{3}\pi  (2 \pi -\sqrt{3})  \omega -23 \pi ^2+24 \pi  \sqrt{3}-9)+ 2 e^{\pi  \omega } (-2 \pi ^2 \omega ^2-12 \pi  \omega +4 \pi ^2-9)\nonumber\\
&-e^{\frac{4 \pi  \omega }{3}}(24 \pi  \sqrt{3}+23 \pi ^2+9) + e^{\frac{5 \pi  \omega }{3}}(24 \pi  \sqrt{3}+23 \pi ^2+9)+ 2 e^{2 \pi  \omega } \left(2 \pi ^2 \omega ^2-12 \pi  \omega -4 \pi ^2+9\right)
\Bigg).
\end{align}
Comparing them we see a clean structure there, which is consistent with the bootstrap.

\section{The unitarity and crossing symmetry}
\label{sec:base-integral}

After the Fourier transform, the  unitarity and crossing   symmetry of the 2d S-matrix $S(\theta)$ connect celestial amplitudes of different orders $f_{n}^{\pm}(\omega)$. The discussion here parallels the original analysis~\cite{Duary:2022onm}, except that there are additional contributions from the bound states in the physical strip. This turns out to contribute additional residues to the results of~\cite{Duary:2022onm}. To keep conciseness, we only show the key formulae below. The interested reader can refer to the original work~\cite{Duary:2022onm} for a pedagogical explanation and the work~\cite{Kapec:2022xjw} for the effect of bound states in the approach coupled with gravity. 

With the $\pm i\epsilon$ regulator, the crossing symmetry becomes $S(\theta+i\epsilon)=S(i\pi-\theta+i\epsilon)$ and its Fourier transform gives
\begin{align}
\label{eq:crossing_retarded_celestial_amp}
\mathcal A^+(\omega) =&\int^{\infty}_{-\infty}d\theta e^{i\omega\theta}S(\theta+i\epsilon)=\int^{\infty}_{-\infty}d\theta e^{i
\omega\theta}S(i\pi-\theta+i\epsilon)=\int^{i\pi+\infty}_{i\pi-\infty}d\theta'e^{i\omega(i\pi-\theta')}S(\theta'+i\epsilon)\nonumber\\
&=e^{-\pi\omega}\left(\mathcal A^+(-\omega)-2\pi i\sum^2_{k=1}\mathrm{Res}\left[e^{-i\omega\theta}S(\theta)\right]_{\theta=i\frac{k\pi}{3}}\right).
\end{align}
Similarly the crossing symmetry  $S(\theta-i\epsilon)=S(i\pi-\theta-i\epsilon)$ acts on the advanced celestial amplitudes as 
\begin{align}
\label{eq:crossing_advanced_celestial_amp}
\mathcal A^-(\omega)&=e^{\pi\omega}\left(\mathcal A^-(-\omega)+2\pi i\sum^2_{k=1}\mathrm{Res}\left[e^{-i\omega\theta}S(\theta)\right]_{\theta=-i\frac{k\pi}{3}}\right)\quad.
\end{align}
This becomes the following relation for the perturbative amplitudes
\begin{align}
\label{eq:crossing_condtion}
f^{\pm}_n(\omega)=e^{\mp\omega\pi}\left(f^{\pm}_n(-\omega)\mp i\sum^2_{k=1}\mathrm{Res}\left[e^{-i\omega\theta}S^{(n)}(\theta)\right]_{\theta=\pm i\frac{k\pi\omega}{3}}\right).
\end{align}
So the crossing symmetry connects the $f^{\pm}_n(\omega)$ with $f^{\pm}_n(-\omega)$ and the residue term is the extra information coming from the bound states. For the 2d S-matrix the crossing symmetry relates the  s-channel with the t-channel scattering  process. This scattering process is sure to be affected by an extra intermediate bound state. On the celestial dual, the crossing symmetry relates physics of the scaling dimension $\omega$ with its inverted dimension $-\omega$ and the effect of bound states simplifies into residues. 

On the other hand, the unitarity $S(\theta+i \epsilon) S(-\theta-i \epsilon)=1$ connects the retarded and the advanced amplitudes after the Fourier transform
\begin{align}
2\pi\delta(\omega)=\int_{-\infty}^{\infty} d \theta e^{i \omega \theta} S(\theta+i \epsilon) S(-\theta-i \epsilon)  =\frac{1}{2 \pi} \int_{-\infty}^{\infty} d \omega^{\prime} \mathcal{A}^{-}\left(\omega^{\prime}\right) \mathcal{A}^{+}\left(\omega+\omega^{\prime}\right).
\end{align}
This is essentially the convolution except a minus sign in the argument. For convenience it is still called the convolution in most circumstances. The relation connects the perturbative amplitudes of different orders as following
\begin{align}
    \label{eq:unitarity_condtion}
&f^+_1(\omega)+f^-_1(-\omega)=0\quad,\notag\\
&f^+_n(\omega)+f^-_n(-\omega)+\sum^{n-1}_{j=1}\int^{\infty}_{-\infty}d\omega'f^+_{n-j}(\omega+\omega')f^-_j(\omega')=0\quad(n\geq 2).
\end{align}

The amplitudes $f^{\pm}_{1,2,3,4}(\omega)$ computed by the Cauchy integration formula are given in~\eqref{eq:perturbative_celestial_amplitude_retarded},~\eqref{eq:perturbative_celestial_amplitude_advanced},~\eqref{eq:f4p} and ~\eqref{eq:f4m}. It is a simple algebra to check that they indeed satisfy the crossing relation~\eqref{eq:crossing_condtion}. 

The difficulty is in the unitarity relation because of the convolution integral. However, the convolution integration turns out to be calculable with the 7 basis of the finite-dimensional linear space~\eqref{eq:7basis}. From it we can see that the convolution only involves $11$ kinds of integrands 
\begin{align}
\label{eq:11basis-convolution}
\int d\omega'f^+_{n-j}(\omega+\omega')f^-_j(\omega') & \propto \sum_{q=0}^{5} \sum_{k=1}^{11} e^{\frac{q \pi\omega}{3}} \int d\omega' \frac{P_{q;n-2}(\omega') e^{\frac{k \pi\omega'}{3}}}{(-1+e^{2\pi(\omega+\omega')})(-1+e^{2\pi\omega'})}
\end{align}
This kind of integral can be done by the Feynman tricks of integration. 

To begin, let's define the base integral with the generating parameter $\alpha$
\begin{align}
\label{eq:I_omega_n_alpha_1}
I(\omega,k,\alpha)&=\int^{\infty}_{-\infty}d\omega'\frac{e^{\frac{k\alpha\pi}{3}\omega'}}{(-1+e^{2\pi(\omega+\omega')})(-1+e^{2\pi\omega'})}, \quad k=1,2,\ldots,11, \nonumber\\
I^{[m]}(\omega,k,\alpha)&=\left(\frac{3}{k\pi}\right)^{m}\frac{\partial^{m} I(\omega,k,a)}{\partial^{m} \alpha}=\int^{\infty}_{-\infty}d\omega'\frac{\omega^{\prime m}e^{\frac{k\alpha\pi}{3}\omega'}}{(-1+e^{2\pi(\omega+\omega')})(-1+e^{2
\pi\omega'})},
\end{align}
with the understanding $I^{[0]}(\omega,k,\alpha)=I(\omega,k,\alpha)$ for the programming purpose. This generating integral covers the $11$ different kinds of integrals of the convolution. Once $I(\omega,k,\alpha)$ is known, this convolution problem would be solved. Let's simplify it as follows 
\begin{align}
\label{eq:I_omega_n_alpha_2}
&I(\omega,k,\alpha)\xlongequal{x=e^{2\pi\omega'}}\int^{\infty}_{0}\frac{dx}{2\pi }\frac{x^{k\alpha/6-1}}{(-1+e^{2\pi\omega}x)(-1+x)} \notag\\ 
&\xlongequal{\text{PF}}\frac{1}{2\pi(e^{2\pi\omega}-1)}\int^\infty_0dx\left(\frac{x^{k\alpha/6-1}}{e^{-2\pi\omega}-x}-\frac{x^{k\alpha/6-1}}{1-x}\right)=\frac{(e^{-2\pi\omega})^{k\alpha/6-1}-1}{2\pi(e^{2\pi\omega}-1)}\int^\infty_0dx\frac{x^{k\alpha/6-1}}{1-x}, 
\end{align}
where a change of variable $y=x e^{2\pi\omega}$ is done in the last step. Now let's split the region as $(0,\infty)=(0,1)\cup(1,\infty)$ and bring $(1,\infty)$ into $(0,1)$ by inverting the variable 
\begin{align}
\label{eq:I_2_2}
\left(\int^1_0dx + \int^\infty_1dx\right)&\frac{x^{k\alpha/6-1}}{1-x}=\int^1_0dx \left(\frac{1-x^{-k\alpha/6}}{1-x}-\frac{1-x^{k\alpha/6-1}}{1-x}\right)\nonumber\\
&=H_{-k\alpha/6} - H_{k\alpha/6-1}=\pi\cot\left(\frac{k\alpha\pi}{6}\right),
\end{align}
where $H_{\alpha}$ is the harmonic number and its reflection formula is used
\begin{align}
H_\alpha=\int_0^1 \frac{1-x^\alpha}{1-x} d x, \quad H_{-\alpha}-H_{\alpha-1}=\pi \cot (\pi \alpha).
\end{align}
So we finally obtain the base integral 
\begin{align}
\label{eq:I_omega_n_alpha_4}
I(\omega,k,\alpha)=\frac{1}{2}\frac{-1+(e^{-2\pi\omega})^{k\alpha/6-1}}{-1+e^{2\pi\omega}}\,\cot\left(\frac{k\alpha\pi}{6}\right).
\end{align}

Note that the analytic continuation of the harmonic number has poles at negative integers, which becomes poles of $\cot(\cdot)$ by the reflection formula.  After the generating parameter is restored to one $\alpha=1$,  one of the $11$ basis of the convolution will overlap with these poles. It is  the $I(\omega,6,1)$, but its  overall factor contributes a zero that exactly cancel this pole of the harmonic number, so $I(\omega,6,1)=-\omega/(-1+e^{2\pi\omega})$ is well behaved.  In this way all the $11$ basis of the convolution are well behaved.
From the expression of $I(\omega,k,\alpha)$~\eqref{eq:I_omega_n_alpha_4}, $I^{[m]}(\omega,k,\alpha)$ can be computed easily. With the help of them, all integrals of the convolution can be done by simple algebra. 

As an example, the convolution between  $f^+_1(\omega)$ and $f^-_1(\omega)$ is evaluated below
\begin{align}
\label{eq:I11}
&\int^{\infty}_{-\infty}d\omega'f^+_1(\omega+\omega')f^-_1(\omega')=\frac{4\pi^2}{9}\Big(2I(\omega,1,1)+\left(-2+e^{\frac{\pi\omega}{3}}\right) I(\omega,2,1)\nonumber\\
&+\left(-4-e^{\frac{\pi\omega}{3}}-e^{\frac{2\pi\omega}{3}}\right)I(\omega,3,1)+\left(-2-2e^{\frac{\pi\omega}{3}}+e^{\frac{2\pi\omega}{3}}-2e^{\pi\omega}\right)I(\omega,4,1)\nonumber\\
&+\left(2-e^{\frac{\pi\omega}{3}}+2e^{\frac{2\pi\omega}{3}}+2e^{\pi\omega}-e^{\frac{4\pi\omega}{3}}\right)I(\omega,5,1)+\left(4+e^{\frac{\pi\omega}{3}}+e^{\frac{2\pi\omega}{3}}+4e^{\pi\omega}+e^{\frac{4\pi\omega}{3}}+e^{\frac{5\pi\omega}{3}}\right)I(\omega,6,1)\nonumber\\
&+\left(2e^{\frac{\pi\omega}{3}}-e^{\frac{2\pi\omega}{3}}+2e^{\pi\omega}+2e^{\frac{4\pi\omega}{3}}-e^{\frac{5\pi\omega}{3}}\right)I(\omega,7,1)+\left(-2e^{\frac{2\pi\omega}{3}}-2e^{\pi\omega}+e^{\frac{4\pi\omega}{3}}-2e^{\frac{5\pi\omega}{3}}\right)I(\omega,8,1)\nonumber\\
&+\left(-4e^{\pi\omega}-e^{\frac{4\pi\omega}{3}}-e^{\frac{5\pi\omega}{3}}\right)I(\omega,9,1)+\left(-2e^{\frac{4\pi\omega}{3}}+e^{\frac{5\pi\omega}{3}}\right)I(\omega,10,1)+2e^{\frac{5\pi\omega}{3}}I(\omega,11,1)\Big)\nonumber\\
&=\frac{4\pi^2(e^{\frac{2\pi\omega}{3}}(4\sqrt{3}-\omega)-4\omega-e^{\frac{\pi\omega}{3}}(4\sqrt{3}+\omega))}{9(-1+e^{\pi\omega})}.
\end{align}
In this manner,  the convolution of amplitudes for the first four orders are computed. For conciseness, we show only the explicit expression for the convolution of the next order as follows
\begin{align}
\label{eq:Int_f2pf1m}
&\int^{\infty}_{-\infty}d\omega'f^+_2(\omega+\omega')f^-_1(\omega')=-\frac{2 \pi ^2 }{27 \sqrt{3} \left(e^{2 \pi  \omega }-1\right)}\Big(8 \sqrt{3} \left(\pi  \left(\omega ^2-3\right)-3 \omega \right) \nonumber\\
&+e^{\frac{\pi  \omega }{3}} \left(2 \pi  \sqrt{3} \omega ^2-6 \sqrt{3} \omega +48 \pi  \omega +69 \pi  \sqrt{3}-72\right)\nonumber\\
&+e^{\frac{2 \pi  \omega }{3}} \left(\pi  \left(-2 \sqrt{3} \omega ^2+48 \omega -69 \sqrt{3}\right)-6 \sqrt{3} \omega +72\right)-8 \sqrt{3} e^{\pi  \omega } \left(\pi  \left(\omega ^2-3\right)+3 \omega \right) \nonumber\\
&-3 e^{\frac{4 \pi  \omega }{3}} \left(2 \sqrt{3} \omega +8 \pi  \omega +23 \pi  \sqrt{3}+24\right)+3 e^{\frac{5 \pi  \omega }{3}} \left(-2 \sqrt{3} \omega -8 \pi  \omega +23 \pi  \sqrt{3}+24\right)\Big).
\end{align}
\begin{align}
\label{eq:Int_f1pf2m}
&\int^{\infty}_{-\infty}d\omega'f^+_1(\omega+\omega')f^-_2(\omega')=\frac{2 \pi ^2 }{27 \sqrt{3} \left(e^{2 \pi  \omega }-1\right)}\Big(8 \sqrt{3} \left(\pi  \left(\omega ^2-3\right)+3 \omega \right) \nonumber\\
&+3 e^{\frac{\pi  \omega }{3}} \left(2 \sqrt{3} \omega +8 \pi  \omega +23 \pi  \sqrt{3}+24\right) +e^{\frac{2 \pi  \omega }{3}} \left(6 \sqrt{3} \omega +24 \pi  \omega -69 \sqrt{3} \pi -72\right)\nonumber\\
&-8 \sqrt{3} e^{\pi  \omega } \left(\pi  \left(\omega ^2-3\right)-3 \omega \right)+e^{\frac{4 \pi  \omega }{3}} \left(-\pi  \left(2 \sqrt{3} \omega ^2+48 \omega +69 \sqrt{3}\right)+6 \sqrt{3} \omega +72\right)\nonumber\\
&+e^{\frac{5 \pi  \omega }{3}} \left(\pi  \left(2 \sqrt{3} \omega ^2-48 \omega +69 \sqrt{3}\right)+6 \sqrt{3} \omega -72\right)\Big).
\end{align}
The detailed computation of these two integrals are given in the appendix~\eqref{eq:I21} and ~\eqref{eq:I12} to exhibit the algorithm more clearly. In this way, we explicitly verified the unitarity condition for the first four orders
\begin{align}
\label{eq:unitarity_condtion-verify}
&f^+_1(\omega)+f^-_1(-\omega)=0,\nonumber\\
&f^+_2(\omega)+f^-_2(-\omega)+\int^{\infty}_{-\infty}d\omega'f^+_{1}(\omega+\omega')f^-_1(\omega')=0,\nonumber\\
&f^+_3(\omega)+f^-_3(-\omega)+\int^\infty_{-\infty}d\omega'\left[f^+_{1}(\omega+\omega')f^-_2(\omega')+f^+_{2}(\omega+\omega')f^-_1(\omega')\right]=0, \nonumber\\   
&f^+_4(\omega)+f^-_4(-\omega)+\int^\infty_{-\infty}d\omega'\left[f^+_{1}(\omega+\omega')f^-_3(\omega')+f^+_{2}(\omega+\omega')f^-_2(\omega')+f^+_{3}(\omega+\omega')f^-_1(\omega')\right]=0.
\end{align}
To keep conciseness, the explicit expression of convolutions of higher orders is not shown in the paper. They can be obtained easily by the bootstrap procedure, as discussed in the next section.

\section{Higher-order terms and bootstrap}
\label{sec:bootstrap}

The true power of the unitariy and crossing symmetry is that they can organize the perturbative  celestial amplitudes into a celestial bootstrap, which for the  Bullough-Dodd model is a practically workable procedure.
The definition of perturbative celestial amplitudes gives 
\begin{align}
\label{eq:Residue0_bootstrap}
f^+_n(\omega)-f^-_n(\omega)+i\, \mathrm{Res}\left[e^{i\omega\theta}S^{(n)}(\theta)\right]_{\theta=0}=0.
\end{align}
Combined with the unitarity~\eqref{eq:unitarity_condtion} and crossing symmetry~\eqref{eq:crossing_condtion}, we get the following bootstrap relation 
\begin{align}
\label{eq:fn+_omega}
&f^+_n(\omega)=\frac{1}{(1+e^{-\pi\omega})}\Bigg[-e^{-\pi\omega} i\,\mathrm{Res}\left[e^{i\omega\theta}S^{(n)}(\theta)\right]_{\theta=0}\notag\\
&+ i\,\sum^2_{k=1}\mathrm{Res}\left[e^{-i\omega\theta}S^{(n)}(\theta)\right]_{\theta=-i\frac{k\pi\omega}{3}}-\sum^{n-1}_{j=1}\int^\infty_{-\infty}d\omega'f^+_{n-j}(\omega+\omega')f^-_j(\omega')\Bigg].
\end{align}
\begin{align}
\label{eq:fn-_omega}
&f^-_n(\omega)=\frac{1}{(1+e^{-\pi\omega})}\Bigg[ i\,\mathrm{Res}\left[e^{i\omega\theta}S^{(n)}(\theta)\right]_{\theta=0}\notag\\
&+ i\,\sum^2_{k=1}\mathrm{Res}\left[e^{-i\omega\theta}S^{(n)}(\theta)\right]_{\theta=-i\frac{k\pi\omega}{3}}-\sum^{n-1}_{j=1}\int^\infty_{-\infty}d\omega'f^+_{n-j}(\omega+\omega')f^-_j(\omega')\Bigg]\quad.
\end{align}
Compared with the bootstrap equation~\cite{Duary:2022onm} of the Sinh-Gordon model, here two extra residues appear due to the bound states of  the Bullough-Dodd model. 

From the general structure of the residues~\eqref{eq:sn-residue}, the three residues at $0,-i\pi/3,-2i\pi/3$ belong to the $7$ kinds of exponential factors. The convolution integral also belongs to the $7$ kinds of exponential factors as in~\eqref{eq:11basis-convolution}.  In practice, the $7$-basis structure of $f_n^{\pm}(\omega)$ is preserved by the celestial bootstrap after transforming the denominator to $(e^{2 \pi  \omega }-1)$. This structure has been verified by boostrapping celestial amplitudes up to $f_{8}^{\pm}(\omega)$. 

Explicitly, the celestial boostrap of the  Bullough-Dodd model has the following algorithm. Suppose the perturbative amplitudes $f_{N}^{\pm}(\omega)$ are wanted. For preparation we need to expand the 2d S-matrix into the N-th order and compute the celestial amplitudes of the first order $f_{1}^{\pm}(\omega)$. Then the following steps will output the target amplitudes $f_{N}^{\pm}(\omega)$. 
\begin{algorithmic}[1]
\Require $\left\{S^{(1)}(\theta), S^{(2)}(\theta),\ldots, S^{(N)}(\theta) \right\},$ $f^{+}_1(\omega), f^{-}_1(\omega)$
\State $n\gets 2$
\While{$n \leq N$}
\State compute residue: $\mathrm{Res}\left[e^{i\omega\theta}S^{(n)}(\theta)\right]_{\theta=0}$, $\mathrm{Res}\left[e^{-i\omega\theta}S^{(n)}(\theta)\right]_{\theta=-i\frac{\pi\omega}{3},-i\frac{2\pi\omega}{3}}$
\State compute $\displaystyle\sum^{n-1}_{j=1}\int^\infty_{-\infty}d\omega'f^+_{n-j}(\omega+\omega')f^-_j(\omega')$ using  $\displaystyle I^{[m]}(\omega,k,\alpha)=(\frac{3}{k\pi})^{m}\frac{\partial^m I(\omega,k,a)}{\partial^m \alpha}$
\State get $f^{+}_n(\omega),f^{-}_n(\omega)$ from the bootstrap formula
\State $n\gets n+1$
\EndWhile
\end{algorithmic}
Using this we have computed the amplitudes upto $f_{8}^{\pm}(\omega)$. The first four orders match the direct computation by the Cauchy integration formula. All of them satisfy the $7$-basis structure~\eqref{eq:7basis}. We do this in a semi-automatic manner. But for programming experts it can be made full-automatic.

\section{Revisit the Sinh-Gordon model}
\label{sec:sinh-gordon}

In the original work~\cite{Duary:2022onm}, the convolution integral of the Sinh-Gordon model is computed at the first order, that is $\int d\omega' f^{+}_1(\omega+\omega')f^{-}_1(\omega')$. Equipped with the methods of this paper, we revisit it further and find that its celestial bootstrap belongs to a linear subspace of that spanned by the Bullough-Dodd model. For conciseness, we  list only the minimal set of equations necessary to explain this, because most equations are already given in~\cite{Duary:2022onm}. 

The exact 2d S-matrix of the Sinh-Gordon model~\cite{Zamolodchikov:1978xm} is 
\begin{align}
    S(\theta)=\frac{\sinh \theta-i \sin \alpha}{\sinh \theta+i \sin \alpha} \quad, \quad \alpha=\frac{\pi b^2}{8 \pi+b^2}, 
\end{align}
whose perturbative series expansion in terms of $b$ has the following structure  
\begin{align}
\label{eq:sn-expansion-SG}
    S^{(n)}(\theta)\propto \mathrm{csch}^{n}(\theta)\Big(\text{upto }\sinh{(n-1)\theta} \text{ or } \cosh{(n-1)\theta}\Big).
\end{align}
As an example, the first three are shown  as follows 
\begin{align}
\label{eq:Series[S]-SG}
&S^{(1)}(\theta)=-\frac{1}{4} i \operatorname{Csch}(\theta)\notag\\
&S^{(2)}(\theta)=-\frac{\operatorname{Csch}(\theta)^2(-i \sinh(\theta)+\pi )}{32 \pi}\nonumber\\
&S^{(3)}(\theta)=\frac{i \operatorname{Csch}(\theta)^3\left(6+11 \pi^2+\left(-6+\pi^2\right) \operatorname{Cosh}(2 \theta)-24 i \sinh (\theta)\right)}{3072 \pi^2},
\end{align}
which is nothing but a reorganize of results of~\cite{Duary:2022onm}.

Following the same method as Section~\ref{sec:amp}, its celestial amplitudes constitutes a linear subspace of dimension-three
\begin{align}
\label{eq:3basis}
   f_{n}^{+}(\omega) =\frac{1}{e^{2 \pi  \omega }-1} \sum_{k=0}^{1} P^{+}_{k;n-1}(\omega) e^{k\pi\omega}, \quad f_{n}^{-}(\omega) =\frac{1}{e^{2 \pi  \omega }-1} \sum_{k=1}^{2} P^{-}_{k;n-1}(\omega) e^{k\pi\omega},
\end{align}
which is exactly the subspace $k=0, 3, 6$ of the  Bullough-Dodd model~\eqref{eq:7basis}. 
Again these relations are  confirmed in hindsight by the celestial bootstrap  and everything appears to be consistent with each other. 
As an example, the first three orders are explicitly shown below
\begin{align}
\label{eq:perturbative_SG_retarded}
&f^+_1=\frac{1-e^{\pi \omega}}{4\left(-1+e^{2 \pi \omega}\right)}, \quad f^+_2=\frac{-1+\pi \omega+e^{\pi \omega}(1+\pi \omega)}{32\left(-1+e^{2 \pi \omega}\right) \pi}\notag\\
&f^+_3=\frac{6-12 \pi \omega+\pi^2\left(2+3 \omega^2\right)-e^{\pi \omega}\left(6+12 \pi \omega+\pi^2\left(2+3 \omega^2\right)\right)}{1536\left(-1+e^{2 \pi \omega}\right) \pi^2}.
\end{align}
Similarly the perturbative advanced celestial amplitude are
\begin{align}
\label{eq:perturbative_SG_advanced}
&f^-_1=\frac{-e^{\pi \omega}+e^{2\pi \omega}}{4\left(-1+e^{2 \pi \omega}\right)}, \quad  f^-_2=\frac{e^{\pi \omega}(1+\pi \omega)+e^{2\pi \omega}(-1+\pi \omega)}{32\left(-1+e^{2 \pi \omega}\right) \pi}\notag\\
&f^-_3=\frac{-e^{\pi \omega}\left(6+12 \pi \omega+\pi^2\left(2+3 \omega^2\right)\right)+e^{2\pi \omega}\left(6-12 \pi \omega+\pi^2\left(2+3 \omega^2\right)\right)}{1536\left(-1+e^{2 \pi \omega}\right) \pi^2}.
\end{align}
Clearly they satisfy the general structure just mentioned. 

Then the convolution integrals can be classified into three different kinds as
\begin{align}
\label{eq:3basis-convolution}
\int d\omega'f^+_{n-j}(\omega+\omega')f^-_j(\omega') & \propto \sum_{q=0}^{1} \sum_{k=1}^{3} e^{k \pi\omega'} \int d\omega' \frac{P_{q;n-2}(\omega') e^{k \pi\omega'}}{(-1+e^{2\pi(\omega+\omega')})(-1+e^{2\pi\omega'})},
\end{align}
which is exactly the subset  $k=3, 6, 9$ of the convolutions of Bullough-Dodd model~\eqref{eq:11basis-convolution}. So they can also be evaluated by the generating function $I^{[m]}(\omega,k,\alpha)$ and the celestial bootstrap goes on similarly as Section~\ref{sec:bootstrap}. We  calculate explicitly using the Cauchy integration formula up to $f_{5}^{\pm}(\omega)$ for  the Sinh-Gordon model and they match results from the celestial bootstrap. In this way, the perturbative celestial amplitudes belongs to the dimension-three subspace of the dimension-seven linear space spanned by amplitudes of the Bullough-Dodd model.

\section{Conclusion}
\label{sec:conclusion}

In this work, we studied the celestial dual of the 2d S-matrix of the integrable Bullough-Dodd model. Its celestial amplitudes constitute a finite-dimensional linear space, where the base-integral is connected with harmonic numbers. The celestial bootstrap becomes an algebraic recursion process. We computed celestial amplitudes directly up to the fourth order. Given the celestial amplitude of the first order, the bootstrap process generates higher order amplitudes and verifies results of the direct computation. 

It is interesting that the linear space spanned by celestial amplitudes of the Bullough-Dodd model has a subspace spanned by amplitudes of the Sinh-Gordon model. These two models are the simplest 2d integrable models. The Sinh-Gordon model has no bound state and the Bullough-Dodd model has the particle behaving as the bound state itself, so none of them has solitons in their spectrum. Intuitively this might be the physical reason why their celestial amplitudes share the same structure. It might be possible that the presence of solitons can change this simple structure. The Sine-Gordon model is a famous one with solitons, whose  2d  S-matrix bootstrap is studied recently in~\cite{Stolbova:2023smk}. It is already a complicated problem at the 2d level, so its celestial bootstrap would be more complicated. Currently we treat this as an open question. 

The analytically-continued harmonic number has poles at  negative integers. One of the base-integral touches the negative integer at $-1$ of the harmonic number, but luckily the overall factors cancel this pole and makes the integral well behaved. This is an interesting phenomenon, and things like this happen frequently in physics. 

The celestial bootstrap in this example makes it practical to compute celestial amplitudes up to arbitrary order. But in the end, it’s a recursive process, not a closed formula. Because the 2d S-matrix is exact, intuitively we expect that the celestial amplitudes should have a closed formula.  We have the consistency among the general structure of the perturbative S-matrix~\eqref{eq:sn-expansion}, its residues~\eqref{eq:sn-residue}, the $7$-basis structure~\eqref{eq:7basis} of $f_{n}^{\pm}(\omega)$ and the $11$-basis structure~\eqref{eq:11basis-convolution} of the convolution. This increases the hope to find such a formula. In the literature of 2d integrable QFTs, closed formula are usually seen by methods of integrability. They might be inspiring to the search of the closed formula of celestial amplitudes. This is an open question. 

We only considered the 2-to-2 scattering in this example. By the factorization property of 2d S-matrix, in principle it might be possible to do the celestial bootstrap for arbitrary n-to-n scattering. Currently we have no clue how to do it practically. The discussion in the appendix of~\cite{Kapec:2022xjw} might be helpful in this direction. This is another open question.

\acknowledgments
We thank Zijian Zhao for helpful discussions. Wei Fan is supported in part by the National Natural Science Foundation of China under Grant No.\ 12105121.

\appendix
\section{Long integrals}

\begin{align}
\label{eq:f4p}
&f^+_4(\omega)=\frac{2 \pi  }{729 \left(e^{2 \pi  \omega }-1\right)}\Big(8 \pi ^3 (3 \omega ^3-18 \omega +20 \sqrt{3})-324 \pi ^2 (\omega ^2-2)+972 \pi  \omega -486 \nonumber\\
&+e^{\frac{\pi  \omega }{3}} (-81 \pi ^2 (2 \omega ^2+8 \sqrt{3} \omega +23)+2 \pi ^3 (6 \omega ^3+36 \sqrt{3} \omega ^2+207 \omega +226 \sqrt{3})+486 \pi  (\omega +2 \sqrt{3})-243)\nonumber\\
&+e^{\frac{2 \pi  \omega }{3}} (-81 \pi ^2 (-2 \omega ^2+8 \sqrt{3} \omega -23)-2 \pi ^3 (-6 \omega ^3+36 \sqrt{3} \omega ^2-207 \omega +226 \sqrt{3})-486 \pi  (2 \sqrt{3}-\omega )+243)\nonumber\\
&+e^{\pi  \omega } (-8 \pi ^3 (-3 \omega ^3+18 \omega +20 \sqrt{3})+324 \pi ^2 (\omega ^2-2)+972 \pi  \omega +486)\nonumber\\
&+e^{\frac{4 \pi  \omega }{3}}(292 \pi ^3 \sqrt{3}+972 \pi  \sqrt{3}+1863 \pi ^2+243) +e^{\frac{5 \pi  \omega }{3}}(-292 \pi ^3 \sqrt{3}-972 \pi  \sqrt{3}-1863 \pi ^2-243) 
\Big)
\end{align}

\begin{align}
\label{eq:f4m}
&f^-_4(\omega)=\frac{2 \pi  }{729 \left(e^{2 \pi  \omega }-1\right)}\Big(e^{2\pi  \omega }(8 \pi ^3 (3 \omega ^3-18 \omega +20 \sqrt{3})-324 \pi ^2 (\omega ^2-2)+972 \pi  \omega -486) \nonumber\\
&+e^{\frac{\pi  \omega }{3}} (-81 \pi ^2 (2 \omega ^2+8 \sqrt{3} \omega +23)+2 \pi ^3 (6 \omega ^3+36 \sqrt{3} \omega ^2+207 \omega +226 \sqrt{3})+486 \pi  (\omega +2 \sqrt{3})-243)\nonumber\\
&+e^{\frac{2 \pi  \omega }{3}} (-81 \pi ^2 (-2 \omega ^2+8 \sqrt{3} \omega -23)-2 \pi ^3 (-6 \omega ^3+36 \sqrt{3} \omega ^2-207 \omega +226 \sqrt{3})-486 \pi  (2 \sqrt{3}-\omega )+243)\nonumber\\
&+e^{\pi  \omega } (-8 \pi ^3 (-3 \omega ^3+18 \omega +20 \sqrt{3})+324 \pi ^2 (\omega ^2-2)+972 \pi  \omega +486)\nonumber\\
&+e^{\frac{4 \pi  \omega }{3}}(292 \pi ^3 \sqrt{3}+972 \pi  \sqrt{3}+1863 \pi ^2+243) +e^{\frac{5 \pi  \omega }{3}}(-292 \pi ^3 \sqrt{3}-972 \pi  \sqrt{3}-1863 \pi ^2-243) 
\Big).
\end{align}

\begin{align}
\label{eq:I21}
&\int^{\infty}_{-\infty}d\omega'f^+_2(\omega+\omega')f^-_1(\omega')=\frac{4\pi^2}{27\sqrt{3}}\Big(\Big(-6\sqrt{3}+4\sqrt{3}\pi\omega\Big)I(\omega,1,1)+4\sqrt{3}\pi I^{[1]}(\omega,1,1)\nonumber\\
&+\Big(6\sqrt{3}-4\sqrt{3}\pi\omega+(-3\sqrt{3}+12\pi+2\sqrt{3}\pi\omega)e^{\frac{\pi\omega}{3}}\Big) I(\omega,2,1)+\Big(-4\sqrt{3}\pi+2\sqrt{3}\pi e^{\frac{\pi\omega}{3}}\Big)I^{[1]}(\omega,2,1)\nonumber\\
&+\Big(12\sqrt{3}-8\sqrt{3}\pi\omega+(3\sqrt{3}-12\pi-2\sqrt{3}\pi\omega)e^{\frac{\pi\omega}{3}}+(3\sqrt{3}-12\pi+2\sqrt{3}\pi\omega)e^{\frac{2\pi\omega}{3}}\Big)I(\omega,3,1)\nonumber\\
&+(-8\sqrt{3}\pi-2\sqrt{3}\pi e^{\frac{\pi\omega}{3}}+2\sqrt{3}\pi e^{\frac{2\pi\omega}{3}})I^{[1]}(\omega,3,1)+\Big(6\sqrt{3}-4\sqrt{3}\pi\omega+(6\sqrt{3}+24\pi-4\sqrt{3}\pi\omega)e^{\frac{\pi\omega}{3}}\nonumber\\
&+(-3\sqrt{3}+12\pi-2\sqrt{3}\pi\omega)e^{\frac{2\pi\omega}{3}}+(6\sqrt{3}+4\sqrt{3}\pi\omega)e^{\pi\omega}\Big)I(\omega,4,1)\nonumber\\
&+\Big(-4\sqrt{3}\pi-4\sqrt{3}\pi e^{\frac{\pi\omega}{3}}-2\sqrt{3}\pi e^{\frac{2\pi\omega}{3}}+4\sqrt{3}\pi e^{\pi\omega}\Big)I^{[1]}(\omega,4,1)\nonumber\\
&+\Big(-6\sqrt{3}+4\sqrt{3}\pi\omega+(3\sqrt{3}-12\pi-2\sqrt{3}\pi\omega)e^{\frac{\pi\omega}{3}}+(-6\sqrt{3}+24\pi-4\sqrt{3}\pi\omega)e^{\frac{2\pi\omega}{3}}\nonumber\\
&+(-6\sqrt{3}-4\sqrt{3}\pi\omega)e^{\pi\omega}+(3\sqrt{3}+12\pi)e^{\frac{4\pi\omega}{3}}\Big)I(\omega,5,1)\nonumber\\
&+\Big(4\sqrt{3}\pi-2\sqrt{3}\pi e^{\frac{\pi\omega}{3}}-4\sqrt{3}\pi e^{\frac{2\pi\omega}{3}}-4\sqrt{3}\pi e^{\pi\omega}\Big)I^{[1]}(\omega,5,1)\nonumber\\
&+\Big(-12\sqrt{3}+8\sqrt{3}\pi\omega+(-3\sqrt{3}+12\pi+2\sqrt{3}\pi\omega)e^{\frac{\pi\omega}{3}}+(-3\sqrt{3}+12\pi-2\sqrt{3}\pi\omega)e^{\frac{2\pi\omega}{3}}\nonumber\\
&+(-12\sqrt{3}-8\sqrt{3}\pi\omega)e^{\pi\omega}+(-3\sqrt{3}-12\pi)e^{\frac{4\pi\omega}{3}}+(-3\sqrt{3}-12\pi)e^{\frac{5\pi\omega}{3}}\Big)I(\omega,6,1)\nonumber\\
&+\Big(8\sqrt{3}\pi+2\sqrt{3}\pi e^{\frac{\pi\omega}{3}}-2\sqrt{3}\pi e^{\frac{2\pi\omega}{3}}-8\sqrt{3}\pi e^{\pi\omega}\Big)I^{[1]}(\omega,6,1)\nonumber\\
&+\Big((-6\sqrt{3}+24\pi+4\sqrt{3}\pi\omega)e^{\frac{\pi\omega}{3}}+(3\sqrt{3}-12\pi+2\sqrt{3}\pi\omega)e^{\frac{2\pi\omega}{3}}+(-6\sqrt{3}-4\sqrt{3}\pi\omega)e^{\pi\omega}\nonumber\\
&+(-6\sqrt{3}-24\pi)e^{\frac{4\pi\omega}{3}}+(3\sqrt{3}+12\pi)e^{\frac{5\pi\omega}{3}}\Big)I(\omega,7,1)\nonumber\\
&+\Big(4\sqrt{3}\pi e^{\frac{\pi\omega}{3}}+2\sqrt{3}\pi e^{\frac{2\pi\omega}{3}}-4\sqrt{3}\pi e^{\pi\omega}\Big)I^{[1]}(\omega,7,1)\nonumber\\
&+\Big((6\sqrt{3}-24\pi+4\sqrt{3}\pi\omega)e^{\frac{2\pi\omega}{3}}+(6\sqrt{3}+4\sqrt{3}\pi\omega)e^{\pi\omega}+(-3\sqrt{3}-12\pi)e^{\frac{4\pi\omega}{3}}\nonumber\\
&+(6\sqrt{3}+24\pi)e^{\frac{5\pi\omega}{3}}\Big)I(\omega,8,1)+\Big(4\sqrt{3}\pi e^{\frac{2\pi\omega}{3}}+4\sqrt{3}\pi e^{\pi\omega}\Big)I^{[1]}(\omega,8,1)\nonumber\\
&+\Big((12\sqrt{3}+8\sqrt{3}\pi\omega)e^{\pi\omega}+(3\sqrt{3}+12\pi)e^{\frac{4\pi\omega}{3}}+(3\sqrt{3}+12\pi)e^{\frac{5\pi\omega}{3}}\Big)I(\omega,9,1)\nonumber\\
&+8\sqrt{3}\pi e^{\pi\omega}I^{[1]}(\omega,9,1)+\Big((6\sqrt{3}+24\pi)e^{\frac{4\pi\omega}{3}}+(-3\sqrt{3}-12\pi)e^{\frac{5\pi\omega}{3}}\Big)I(\omega,10,1)\nonumber\\
&+(-6\sqrt{3}-24\pi)e^{\frac{5\pi\omega}{3}}I(\omega,11,1)\Big)\nonumber\\
&=-\frac{2 \pi ^2 }{27 \sqrt{3} \left(e^{2 \pi  \omega }-1\right)}\Big(e^{\frac{\pi  \omega }{3}} \left(2 \pi  \sqrt{3} \omega ^2-6 \sqrt{3} \omega +48 \pi  \omega +69 \pi  \sqrt{3}-72\right)\nonumber\\
&-8 \sqrt{3} e^{\pi  \omega } \left(\pi  \left(\omega ^2-3\right)+3 \omega \right)+e^{\frac{2 \pi  \omega }{3}} \left(\pi  \left(-2 \sqrt{3} \omega ^2+48 \omega -69 \sqrt{3}\right)-6 \sqrt{3} \omega +72\right)\nonumber\\
&+8 \sqrt{3} \left(\pi  \left(\omega ^2-3\right)-3 \omega \right)-3 e^{\frac{4 \pi  \omega }{3}} \left(2 \sqrt{3} \omega +8 \pi  \omega +23 \pi  \sqrt{3}+24\right)\nonumber\\
&+3 e^{\frac{5 \pi  \omega }{3}} \left(-2 \sqrt{3} \omega -8 \pi  \omega +23 \pi  \sqrt{3}+24\right)\Big)\quad.
\end{align}

\begin{align}
\label{eq:I12}
&\int^{\infty}_{-\infty}d\omega'f^+_1(\omega+\omega')f^-_2(\omega')=\frac{4\pi^2}{27\sqrt{3}}\Big(\Big(-6\sqrt{3}+24\pi\Big)I(\omega,1,1)+4\sqrt{3}\pi I^{[1]}(\omega,1,1)\nonumber\\
&+\Big(6\sqrt{3}-24\pi+(-3\sqrt{3}+12\pi)e^{\frac{\pi\omega}{3}}\Big) I(\omega,2,1)+\Big(4\sqrt{3}\pi+2\sqrt{3}\pi e^{\frac{\pi\omega}{3}}\Big)I^{[1]}(\omega,2,1)\nonumber\\
&+\Big(12\sqrt{3}+(3\sqrt{3}-12\pi)e^{\frac{\pi\omega}{3}}+(3\sqrt{3}-12\pi)e^{\frac{2\pi\omega}{3}}\Big)I(\omega,3,1)\nonumber\\
&+(8\sqrt{3}\pi+2\sqrt{3}\pi e^{\frac{\pi\omega}{3}}-2\sqrt{3}\pi e^{\frac{2\pi\omega}{3}})I^{[1]}(\omega,3,1)+\Big(6\sqrt{3}+24\pi+6\sqrt{3}e^{\frac{\pi\omega}{3}}\nonumber\\
&+(-3\sqrt{3}+12\pi)e^{\frac{2\pi\omega}{3}}+(6\sqrt{3}-24\pi)e^{\pi\omega}\Big)I(\omega,4,1)\nonumber\\
&+\Big(4\sqrt{3}\pi e^{\frac{\pi\omega}{3}}-2\sqrt{3}\pi e^{\frac{2\pi\omega}{3}}-4\sqrt{3}\pi e^{\pi\omega}\Big)I^{[1]}(\omega,4,1)\nonumber\\
&+\Big(-6\sqrt{3}-24\pi+(3\sqrt{3}+12\pi)e^{\frac{\pi\omega}{3}}-6\sqrt{3}e^{\frac{2\pi\omega}{3}}\nonumber\\
&+(-6\sqrt{3}+24\pi)e^{\pi\omega}+(3\sqrt{3}-12\pi)e^{\frac{4\pi\omega}{3}}\Big)I(\omega,5,1)\nonumber\\
&+\Big(-4\sqrt{3}\pi e^{\frac{2\pi\omega}{3}}-4\sqrt{3}\pi e^{\pi\omega}-2\sqrt{3}\pi e^{\frac{4\pi\omega}{3}}\Big)I^{[1]}(\omega,5,1)\nonumber\\
&+\Big(-12\sqrt{3}+(-3\sqrt{3}-12\pi)e^{\frac{\pi\omega}{3}}+(-3\sqrt{3}-12\pi)e^{\frac{2\pi\omega}{3}}\nonumber\\
&-12\sqrt{3}e^{\pi\omega}+(-3\sqrt{3}+12\pi)e^{\frac{4\pi\omega}{3}}+(-3\sqrt{3}+12\pi)e^{\frac{5\pi\omega}{3}}\Big)I(\omega,6,1)\nonumber\\
&+\Big(8\sqrt{3}\pi-8\sqrt{3}\pi e^{\pi\omega}-2\sqrt{3}\pi e^{\frac{4\pi\omega}{3}}+2\sqrt{3}\pi e^{\frac{5\pi\omega}{3}}\Big)I^{[1]}(\omega,6,1)\nonumber\\
&+\Big(-6\sqrt{3}e^{\frac{\pi\omega}{3}}+(3\sqrt{3}+12\pi)e^{\frac{2\pi\omega}{3}}+(-6\sqrt{3}-24\pi)e^{\pi\omega}\nonumber\\
&-6\sqrt{3}e^{\frac{4\pi\omega}{3}}+(3\sqrt{3}-12\pi)e^{\frac{5\pi\omega}{3}}\Big)I(\omega,7,1)\nonumber\\
&+\Big(4\sqrt{3}\pi e^{\frac{\pi\omega}{3}}-4\sqrt{3}\pi e^{\frac{4\pi\omega}{3}}+2\sqrt{3}\pi e^{\frac{5\pi\omega}{3}}\Big)I^{[1]}(\omega,7,1)\nonumber\\
&+\Big(6\sqrt{3}e^{\frac{2\pi\omega}{3}}+(6\sqrt{3}+24\pi)e^{\pi\omega}+(-3\sqrt{3}-12\pi)e^{\frac{4\pi\omega}{3}}\nonumber\\
&+6\sqrt{3}e^{\frac{5\pi\omega}{3}}\Big)I(\omega,8,1)+\Big(-4\sqrt{3}\pi e^{\frac{2\pi\omega}{3}}+4\sqrt{3}\pi e^{\frac{5\pi\omega}{3}}\Big)I^{[1]}(\omega,8,1)\nonumber\\
&+\Big(12\sqrt{3}e^{\pi\omega}+(3\sqrt{3}+12\pi)e^{\frac{4\pi\omega}{3}}+(3\sqrt{3}+12\pi)e^{\frac{5\pi\omega}{3}}\Big)I(\omega,9,1)\nonumber\\
&-8\sqrt{3}\pi e^{\pi\omega}I^{[1]}(\omega,9,1)+\Big(6\sqrt{3}e^{\frac{4\pi\omega}{3}}+(-3\sqrt{3}-12\pi)e^{\frac{5\pi\omega}{3}}\Big)I(\omega,10,1)\nonumber\\
&-4\sqrt{3}\pi e^{\frac{4\pi\omega}{3}}I^{[1]}(\omega,10,1)-6\sqrt{3}e^{\frac{5\pi\omega}{3}}I(\omega,11,1)+4\sqrt{3}\pi e^{\frac{5\pi\omega}{3}}I^{[1]}(\omega,11,1)\Big)\nonumber\\
&=\frac{2 \pi ^2 }{27 \sqrt{3} \left(e^{2 \pi  \omega }-1\right)}\Big(-8 \sqrt{3} e^{\pi  \omega } \left(\pi  \left(\omega ^2-3\right)-3 \omega \right)+8 \sqrt{3} \left(\pi  \left(\omega ^2-3\right)+3 \omega \right)\nonumber\\
&+e^{\frac{4 \pi  \omega }{3}} \left(-\pi  \left(2 \sqrt{3} \omega ^2+48 \omega +69 \sqrt{3}\right)+6 \sqrt{3} \omega +72\right)\nonumber\\
&+e^{\frac{5 \pi  \omega }{3}} \left(\pi  \left(2 \sqrt{3} \omega ^2-48 \omega +69 \sqrt{3}\right)+6 \sqrt{3} \omega -72\right)\nonumber\\
&+3 e^{\frac{\pi  \omega }{3}} \left(2 \sqrt{3} \omega +8 \pi  \omega +23 \pi  \sqrt{3}+24\right)+e^{\frac{2 \pi  \omega }{3}} \left(6 \sqrt{3} \omega +24 \pi  \omega -69 \sqrt{3} \pi -72\right)\Big)\quad.
\end{align}

\bibliographystyle{JHEP}
\bibliography{2d_CCFT}

\end{document}